\documentclass[12pt]{article} 

\newenvironment{pf}{\noindent{\bf Proof:}}{\newline\kbn}

\usepackage{amssymb,theorem}

\newtheorem{theo}{Theorem}

\newtheorem{lemma}[theo]{Lemma}
\newtheorem{prop}[theo]{Proposition}

\newcommand{\Einsop}{\leavevmode{\rm 1\mkern  -4.4mu l}}
\newcommand{\Seins}{{\mathsf{S}^1}}

\newcommand{\Diff}{{\mbox{\it Diff}}}
\newcommand{\PSL}{{\mbox{\it PSL}}}

\newcommand{\betrag}[1]{{\left| #1 \right|}} 
 
\newcommand{\klammer}[1]{{\left( #1 \right)}}

\newcommand{\lok}[1]{{\mathcal #1}}
\newcommand{\Hilb}[1]{{\mathcal #1}}
\newcommand{\Name}[1]{{#1}} 
\newcommand{\dopp}[1]{{\mathbb #1}}
\newcommand{\symb}[1]{{\mathbf #1}}
\newcommand{\Geb}[1]{{\mathcal #1}}

\newcommand{\qed}{}
\newcommand{\kbn}{$\square$}

\begin{document}

\title{Implementation of conformal covariance by diffeomorphism
  symmetry} 
\author{Claudio D'Antoni\footnote{Dipartimento di Matematica, Universit\`a di
  Roma ``Tor Vergata''}, Klaus
  Fredenhagen\footnote{II.\ Inst.\ Theor. Physik, Universit\"at \ Hamburg},
  S\o{}ren K\"oster\footnote{Inst.\ f.\ Theor.\ Physik, Universit\"at
  G\"ottingen}} 

\date{} 

\maketitle
\begin{abstract} 
Every 
locally normal 
representation of a local chiral conformal quantum theory is
covariant with respect to global conformal transformations, if this
theory is diffeomorphism covariant in its vacuum
representation. 

The  unitary, strongly
continuous representation implementing conformal symmetry is
constructed; it consists of operators which are inner in 
a global sense for the representation of the quantum theory. 
The construction
is independent of positivity of energy and applies to all locally normal
representations irrespective of their statistical dimensions (index).  
\\[1mm]
AMS Subject classification (2000): 81T40, 81T05, 81R10 
\end{abstract}

\section{Introduction}
\label{sec:intro}

The conformal group in $1+1$ dimensions is an infinite
dimensional diffeomorphism group. Many interesting models exhibit this
symmetry and typically these models factorise into their chiral parts,
each of which depends on one light-cone coordinate only. 
In this short letter we prove 
that all 
locally normal representations of  chiral conformal nets which exhibit
diffeomorphism symmetry in their vacuum representation admit an
implementation of global 
conformal transformations. 

This automatic implementability is interesting since for a class of conformal
models with less symmetry the existence of non-covariant, locally normal
representations has been established by \Name{Guido,
  Longo, Wiesbrock} \cite{GLW98}. Moreover, our method applies in a very
general setting: neither the  representation theory of the
diffeomorphism group, the index of the respective locally normal
representation or even positivity of energy concern our approach at all.
 
Diffeomorphism covariance of a chiral conformal net $\lok{A}$ (see e.g.
\cite{GL96} for general properties)  
means that there is a strongly continuous map $U_0$ from
the group of orientation preserving diffeomorphisms of the circle,
$\Diff_+(\Seins)$, into the unitaries on $\Hilb{H}_0$, the
representation space of the vacuum representation of $\lok{A}$,
implementing a geometric automorphic action
$\alpha$ of $\Diff_+(\Seins)$: $$U_0(\phi) \lok{A}(I)
U_0(\phi)^* \equiv \alpha_\phi(\lok{A}(I)) = \lok{A}(\phi(I))\, , \,\, 
I\Subset\Seins\, , \,\, \phi\in \Diff_+(\Seins)\,\,.$$ The
localisation regions  are {\em proper}, i.e.  open and non-dense, intervals $I$,
denoted $I\Subset\Seins$, whose causal complements are their open
complements, $I'$, in $\Seins$. The subgroup $\Diff_I(\Seins)$ of
diffeomorphisms localised in $I$ consists, by definition, of elements
$\phi\in\Diff_+(\Seins)$ which act trivially on $I'$. For
$\phi\in\Diff_I(\Seins)$ the adjoint action of $U_0(\phi)$ is to
implement the trivial automorphism of $\lok{A}(I')$ and hence it is,
by \Name{Haag} duality of $\lok{A}$, a local observable, i.e.
$U_0(\phi)\in \lok{A}(I)$.

$U_0$ defines a
ray representation, as the cocycles
$U_0(\phi_1)U_0(\phi_2)U_0(\phi_1\phi_2)^*$ commute with
$\lok{A}$ and $\lok{A}$ is irreducible. With results of \Name{Carpi}
\cite{sC03} this shows that $U_0$ corresponds to a definite value of the
central charge of the \Name{Virasoro} algebra. We require
$U_0(id)=\Einsop$ and 
$\alpha\!\restriction\!{\PSL(2,\dopp{R})}$ to be identical to the global conformal
covariance of $\lok{A}$.
In models having a stress-energy tensor, the restricted
representation $U_0\!\restriction\!{\PSL(2,\dopp{R})}$  is in fact a
representation of $\PSL(2,\dopp{R})$. The
further analysis does not require the answer to the cohomological question
whether this may be achieved always by a proper choice of phases for
$U_0$.

We deal with a locally normal representation $\pi$ of
$\lok{A}$, 
i.e. a family of normal 
representations $\pi_I$ of the local algebras $\lok{A}(I)$ by bounded
operators on a \Name{Hilbert} space $\Hilb{H}_\pi$, which is
required to be consistent with isotony: $I\subset J \Rightarrow
\pi_J\!\restriction\!{\lok{A}(I)}= \pi_I$; given this condition is
fulfilled, we say that the local representations $\pi_I$ are {\em
compatible}. 

By local normality, the maps $\pi\circ U_0\restriction
\Diff_I(\Seins)$ define unitary, strongly
continuous projective representations of the respective local
diffeomorphism subgroup with cocycles which are phases, since the
local algebras are factors. We will use the presence of these
representations in order to construct an implementing
ray representation of the subgroup of global conformal transformations
(\Name{Moebius} group $\PSL(2,\dopp{R})$). We begin with a clarification
on the relation of general diffeomorphisms to the representation $\pi$
induced by $U_0$. 

We introduce the {\em universal $C^*$-algebra} $\lok{A}_{uni}$ generated by
the local algebras of $\lok{A}$. The properties of $\lok{A}_{uni}$ are
summarised in 
\begin{prop} \cite{FRS92,GL92} There is a unique $C^*$-algebra $\lok{A}_{uni}$
such that 
\begin{enumerate}
\item For all $I\Subset\Seins$ there exist injective, compatible
embeddings $\iota_I: \lok{A}(I) \rightarrow \lok{A}_{uni}$ and
$\lok{A}_{uni}$ is generated by its subalgebras $\iota_I(\lok{A}(I))$. 
\item For any compatible family of representations $\{\pi_I\}$  there
exists a unique representation $\pi$ of $\lok{A}_{uni}$ by bounded
operators on $\Hilb{H}_\pi$ such that $\pi\circ\iota_I = \pi_I$. 

Moreover, every representation $\pi$ of $\lok{A}_{uni}$ restricts to a
representation of $\lok{A}$. The vacuum representation $\pi_0$ of
$\lok{A}_{uni}$ corresponds to the identity (defining) representation
of $\lok{A}$ on $\Hilb{H}_0$: $\pi_0\restriction \lok{A}(I) = id
\restriction \lok{A}(I)$. 
\end{enumerate}
\end{prop}

The action $\alpha$ of $\Diff_+(\Seins)$ on the net $\lok{A}$ can be
extended to an action by automorphisms of $\lok{A}_{uni}$ through
$\alpha_\phi\circ \iota_I := \iota_{\phi I}\circ
Ad_{U_0(\phi)}\restriction \lok{A}(I)$. Our next goal is to establish
that this action is inner. For this purpose we prove
\begin{prop}
Let $\phi\in\Diff_I(\Seins)$, then for each $J\Subset\Seins$ we have
$\alpha_\phi\restriction \iota_J(\lok{A}(J)) =
Ad_{\iota_{I}(U_0(\phi))} \restriction \iota_J(\lok{A}(J))$.  
Each $\alpha_\phi$, $\phi\in\Diff_+(\Seins)$,  possesses an implementation by unitary elements of
$\lok{A}_{uni}$.
\end{prop}

\begin{pf}
If there is a proper interval $\hat{J}\supset I\cup J$, the statement is
obvious. If $\overline{I\cup J}=\Seins$, we choose a covering
$\{I_i\}_{i=1,2,3}$ of $I$ such that $I_i\Subset\Seins$, 
$\overline{I_3}\subset J'$ and $I_{1,2}\cup J$ both are contained in some
proper interval. By  Lemma \ref{locmap} we find a 
factorisation $\phi = \prod_{i=1}^3\phi_i$, $\phi_i\in \Diff_{I_i}(\Seins)$,
if $\phi$  is contained in the neighbourhood $ \Geb{U}_\varepsilon 
\subset \Diff_I(\Seins)$
defined in the Lemma , for  $\varepsilon$ 
sufficiently small .

The two operators $\iota_I(U_0(\phi))$ and
$\prod_{i=1}^3\iota_{I_i}(U_0(\phi_i))$ coincide up to a scalar multiple of
$\Einsop=\iota_I(\Einsop_{\lok{A(I)}})$. Thus we have:
$$Ad_{\iota_I(U_0(\phi))}\restriction \iota_J(\lok{A}(J)) = \prod_{i=1}^3
Ad_{\iota_{I_i}(U_0(\phi_i))}\restriction \iota_J(\lok{A}(J))\,\,.$$  
This proves the statement for   $\phi\in
\Geb{U}_\varepsilon\cap\Diff_I(\Seins)$. 

Now let $\phi$ be an arbitrary diffeomorphism localised in $I$. We will
factorise $\phi$  into a product of diffeomorphisms such that the above
applies. Let $\varphi$ denote the periodic diffeomorphism of $\dopp{R}$ which
corresponds to $\phi$ (cf proof of lemma \ref{locmap}). Define
$\varphi_\lambda(x) = x + \lambda(\varphi(x)-x)$, $x\in\dopp{R}$, and denote
the corresponding element in $\Diff_I(\Seins)$ by $\phi_\lambda$. 

With the covering $\{I_i\}_{i=1,2,3}$ above there is a  $\delta>0$ such
that for all $\lambda\in [0,1]$ the diffeomorphism
$\phi_{\lambda+\delta}\circ \phi_\lambda^{-1}\in\Geb{U}_\varepsilon$.
Then we can represent $\phi$ as a product
$\phi=\prod_{k=0}^{n-1} (\phi_{k+1/n}\phi_{k/n}^{-1})$,

 for $n$ large enough ,where  each factor and hence $\phi$ 
 satisfies the Proposition.  

Since $\Diff_+(\Seins)$ is a simple group (theorem of
\Name{Epstein, Herman, Thurston}, cf \cite{jM83}) each
$\phi\in\Diff_+(\Seins)$ has a presentation by a finite product of
localised diffeomorphisms. One may take this presentation and the results
proved so far in order to obtain the desired implementation, which completes
the proof.
\qed\end{pf}

 {\em Remark:}  The implementation of a diffeomorphism $\phi$ by an element of
   $\lok{A}_{uni}$ is not unique, in general: two
implementers of $ \alpha_\phi$ may differ by an element from the
centre of $\lok{A}_{uni}$. 

In a representation $\pi$ of
$\lok{A}_{uni}$ the implementers form a projective, unitary
representation of $\Diff_+(\Seins)$ with cocycle in the centre of
$\pi(\lok{A}_{uni})$ and which implements the automorphic action
$\alpha$; if $\pi$ is a factorial representation, the
cocycles are automatically phases.
One would like to derive a {\em genuine} ray representation of
$\Diff_+(\Seins)$ with cocycle which is given by {\em phases} for any general
representation $\pi$, but we only
know a way to do this for the subgroup $\PSL(2,\dopp{R})$ of global conformal
transformations. 

To this end we analyse the cocycle of
the implementation of $\alpha$ and therefore we look for a definite choice of
the implementing unitaries for diffeomorphisms close to the identity:
For elements $\phi$ of a suitable neighbourhood $\Geb{U}_\varepsilon$
of the identity the results of lemma \ref{locmap} (appendix) allow us
to choose a fixed covering $\{I_i\}_{i\in\dopp{Z}_m}$ of $\Seins$ by
proper intervals and a fixed set of localisation maps $\Xi_i:
\Geb{U}_\varepsilon \rightarrow \Diff_{I_i}(\Seins)$ such that
$u_\pi(\phi) := \prod_{i=1}^m \pi\circ U_0\circ\Xi_i (\phi)$
 defines a strongly continuous, unitary and unital map. The adjoint
action of $u_\pi$ induces an implementation of $\alpha$ in $\pi$, which
we will use in the following section.


\section{Obtaining the implementation}
\label{sec:constimpl}

We will now restrict our attention to the subgroup of global conformal
transformations, $\PSL(2,\dopp{R})$,
and construct a unitary, strongly continuous representation of its
universal covering group 
${\PSL(2,\dopp{R})}^\sim$ from
$u_\pi\!\restriction\!{\Geb{U}_\varepsilon\cap \PSL(2,\dopp{R})}$ as
defined at the end of the previous section. This
representation will implement the automorphic 
action $\alpha$ of $\PSL(2,\dopp{R})$ on $\lok{A}$ in the
representation $\pi$ and will be inner in the global sense, i.e. it
will be contained in the \Name{von Neumann} algebra of {\em global
  observables},
$\pi(\lok{A}):=\bigvee_{I\Subset\Seins}\pi_I(\lok{A}(I))$. 

Let us begin with a closer look at the group $\PSL(2,\dopp{R})$
itself. We use the
symbol $T$ for the one-parameter group of translations, $S$ for
the special conformal transformations, $D$ for the scale
transformations (dilatations) and $R$ for rotations. We
choose parameters for the rotations such 
that the rotation group $R$ is naturally isomorphic to
$\dopp{R}/2\pi\dopp{Z}$. 

We can write every $g\in \PSL(2,\dopp{R})$ in the form 
$g = {T}(p_g) {D}(\tau_g) {R}(t_g)$, where each term 
depends continuously on $g$ (\Name{Iwasawa} decomposition, \cite{FG93},
appendix I). In fact, any $g\in \PSL(2,\dopp{R})$ may
be written as a 
product of four translations and four special conformal
transformations, each single of them depending continuously on
$g$, if one uses the identities:
\begin{eqnarray}
  {D}(\tau) &=& {S}(-(e^{\frac{\tau}{2}}-1)e^{-\frac{\tau}{2}})\,
  {T}(1)\, {S}(e^{\frac{\tau}{2}}-1)\,
  {T}(-e^{-\frac{\tau}{2}})\label{eq:stiwa1}
  \label{eq:Dtsdiff} \,\, ,\\  
{R}(2t) &=& {S}((-1+\cos t)(\sin t)^{-1}) \,
{T}(\sin t) \, {S}((-1+\cos t)(\sin t)^{-1})\,\, .
\label{eq:stiwa2}\label{eq:Rtsdiff}
\end{eqnarray}

According to lemma \ref{locmap} (appendix), there are continuous, identity preserving localisation maps $\Xi_j$,
$j=1,..,m$, which map a neighbourhood of the identity,
$\Geb{U}_\varepsilon\subset\Diff_+(\Seins)$, into groups of localised
diffeomorphisms such that we have $\prod_{j=1}^m \Xi_j(\phi) = \phi$,
$\phi\in \Geb{U}_\varepsilon$.
If we specialise to translations, this means that
there is an open interval $I_\varepsilon$ containing  $0$ for which
the mapping $t\mapsto \prod_j u_\pi (\Xi_j(T(t)))$ is unital
and strongly continuous. 
 We extend this map to all
of $\dopp{R}$ through a choice of a $\tau\in I_\varepsilon$, $\tau>0$,
 defining $n_t\in\dopp{Z}$ by its properties $t= n_t \tau+ (t-n_t
\tau)$, $t-n_t 
\tau \in [0,\tau[$, and setting 
$$\pi^\lok{A}(T(t)):= \klammer{\prod_j u_\pi (\Xi_j(T(\tau)))}^{n_t} \prod_j u_\pi
(\Xi_j(T(t-n_t\tau)))\,\,.$$
One can easily check that this is indeed a strongly
continuous map into the unitaries of $B(\Hilb{H}_\pi)$ by recognising
that the mappings involved are continuous and unital ($\pi(\Einsop)=
\Einsop$, $\Xi_i(id)=id$). 

This procedure applies to the special conformal transformations as
 well, and we may use the result, the \Name{Iwasawa} decomposition
 and (\ref{eq:Dtsdiff}), (\ref{eq:Rtsdiff}) to define for each
${g}\in\PSL(2,\dopp{R})$: 
\begin{equation}
  \label{eq:pib}
  \pi^\lok{A}({g}) := \prod_{i=1}^{4}
  T^{\pi(\lok{A})}(t^{(i)}_{{g}}) S^{\pi(\lok{A})}(n^{(i)}_{{g}})\,
  ,\quad {g}\in\PSL(2,\dopp{R}) \,\, .
\end{equation}
We have $\pi^\lok{A}(id)=\Einsop$.
The following Lemma asserts that the
$\pi^\lok{A}({g})$ define an  inner-implementing
representation up to a cocycle in the centre of $\pi(\lok{A})$. To this end
we define operators sensitive to 
the violation of the group multiplication law:
$ z^\lok{A}({{g}},{{h}}):= \pi^\lok{A}({{g}}) \pi^\lok{A}({{h}})
\pi^\lok{A}({{g}}{{h}})^*,\, 
{{g}}, {{h}}\in\PSL(2,\dopp{R})$. 

\begin{lemma}\label{lem:words}
  $\pi^\lok{A}: g\mapsto \pi^\lok{A}(g)$ defines a strongly continuous
  mapping with unitary values in $\pi(\lok{A})$. The adjoint action of
  $\pi^\lok{A}(g)$, $g\in \PSL(2,\dopp{R})$, on $\pi(\lok{A})$ implements the
  automorphism $\alpha_g$. $z^\lok{A}: (g,h)\mapsto z^\lok{A}(g,h)$
  defines a strongly continuous 2-cocycle with unitary values in
  $\pi(\lok{A})'\cap\pi(\lok{A})$.
\end{lemma}

\begin{pf} 
Unitarity is obvious. Strong continuity follows since we multiply
continuous functions. The
implementing property of the $\pi^\lok{A}(g)$ follows immediately by the
decomposition $g=\prod_{i=1}^4 T(t^i_g) S(s^i_g)$, the subsequent
decomposition 
of these into products of localised diffeomorphisms, the definition of
$\pi^\lok{A}(g)$ and the implementation property of the (generalised)
ray representation $u_\pi$ of $\Diff_+(\Seins)$. At
this point all properties of $z^\lok{A}$ follow immediately from
its definition. 
\qed\end{pf}

We write the
abelian \Name{von \-Neu\-mann} algebra generated by the cocycle operators
$z^\lok{A}(g,h)$ as follows:
$
  \lok{Z}^\lok{A} \equiv \{z^\lok{A}(g,h),
      z^\lok{A}(g,h)^* | g,h \in \PSL(2,\dopp{R})\}''
$. Obviously $\lok{Z}^\lok{A}$ is contained in the centre of $\pi(\lok{A})$.
Now we are prepared to realise the construction itself:
\begin{lemma}\label{lem:constr}
  For every $\tilde{g}\in {\PSL(2,\dopp{R})}^\sim$ there exists a
  unitary operator 
  $z^\lok{A}(\tilde{g})\in\lok{Z}^\lok{A}$ such that
  \begin{equation}
    \label{eq:defUA}
    U_\pi(\tilde{g}):= z^\lok{A}(\tilde{g})\pi^\lok{A}(\symb{p}(\tilde{g}))
  \end{equation}
defines a unitary, strongly continuous
representation, whose adjoint action
implements the automorphic action
$\alpha\circ\symb{p}$ on $\pi(\lok{A})$; $\symb{p}$ denotes
the covering projection from ${\PSL(2,\dopp{R})}^\sim$ onto
${\PSL(2,\dopp{R})}$. 
\end{lemma}

\begin{pf} As 
$ \lok{Z}^\lok{A}\subset \pi(\lok{A})\cap \pi(\lok{A})'$ we may apply the direct
integral decomposition (cf e.g. \cite{KR86}, chapter 14). This yields
a decomposition of $\Hilb{H}_\pi$ as a direct integral of \Name{Hilbert}
spaces $\Hilb{H}_x$ and it implies: the action of 
$z^\lok{A}(g,h)$ on $\Hilb{H}_x$, denoted by $z^\lok{A}(g,h)(x)$, is a
multiple of the identity $\Einsop_x$ and thereby defines for
almost every $x$ a  continuous
2-cocycle $\omega(f,g)_x\in\Seins\subset\dopp{C}$. The action of the operators $\pi^\lok{A}(g)$ on
$\Hilb{H}_x$, denoted by $\pi^\lok{A}(g)(x)$, defines for almost
every $x$ a unitary, strongly continuous, projective
representation of ${\PSL(2,\dopp{R})}$, cf \cite{cM76}.

For \Name{Lie} groups with a simple \Name{Lie} algebra the lifting
criterion is valid \cite{dS68}. This ensures for almost every $x$ the
existence of continuous phases $\omega(\tilde{g})(x)$, $\tilde{g}\in
{\PSL(2,\dopp{R})}^\sim$,  such that
$\omega(\tilde{g})(x) \pi^\lok{A}(\symb{p}(\tilde{g}))(x)$ defines a
representation of 
${\PSL(2,\dopp{R})}^\sim$. Integrating $\omega(\tilde{g})(x)$ over all
$x$ yields a 
unitary $z^\lok{A}(\tilde{g})\in \lok{Z}^\lok{A}$, depending strongly
continuously on 
$\tilde{g}$. Integrating the 
$\omega(\tilde{g})(x)\pi^\lok{A}(\symb{p}(\tilde{g}))(x)$ yields a
unitary, strongly continuous 
representation $U_\pi$ satisfying equation
(\ref{eq:defUA}). $U_\pi(\tilde{g})$ is an element of $\pi(\lok{A})$ for every
$\tilde{g}$ and implements $\alpha_{\symb{p}(\tilde{g})}$ by its
adjoint action due to Lemma \ref{lem:words}.
\qed\end{pf}

The outcome of the construction presented above proves our main
result; 
its
uniqueness statement is a simple consequence
of the fact that 
${\PSL(2,\dopp{R})}^\sim$ is a perfect group (cf \cite{sK02}, prop. 2):
\begin{theo}\label{covdiffrep}
  Let $\lok{A}$ be a chiral conformal, diffeomorphism covariant
  theory. Then any locally normal representation 
  $\pi$ of $\lok{A}$ is covariant with respect to the automorphic
  action of $\PSL(2,\dopp{R})$. The implementing representation may be
  chosen to be the unique globally $\pi(\lok{A})$-inner, implementing
  representation $U_\pi$ of ${\PSL(2,\dopp{R})}^\sim$.
\end{theo}

If there are  diffeomorphism covariant theories which possess locally normal
representations violating positivity of 
energy, the construction of the inner-\-implementing representation given here
applies even in cases in which the \Name{Bor\-chers-\-Sugawara} construction
\cite{sK02} cannot be used, as the latter depends on the existence of 
implementations of translations and special conformal transformations which
have positive energy. For representations with finite 
  statistical dimension the spectrum condition is always fulfilled
  because of the theorem we have just
  derived and results of \cite{BCL98}. For infinite index
  representations, of which examples are known (cf \cite{sC02}) there exists a
  criterion  
  for strongly additive theories, which  was given in \cite{BCL98}, too.
 In presence
of the spectrum condition both constructions agree by uniqueness. 

\Name{Yngvason} \cite{jY94} discussed conformally covariant
derivatives of the $U(1)$-current as interesting examples of chiral
conformal theories.
It is straightforward to see that the first conformally covariant
derivative had to exhibit a  diffeomorphism symmetry if
it contained a stress-energy tensor (details in \cite{sK03d}). \Name{Guido,
  Longo, Wiesbrock} studied locally normal representations of
this model \cite{GLW98} and found representations which manifestly do
not admit an implementation of global conformal symmetry. As stated in
 \cite{GLW98} this excludes the presence of a stress-energy tensor
and diffeomorphism symmetry for this model as a consequence of theorem
\ref{covdiffrep}. Presence of a stress-energy tensor may be excluded
directly as shown in \cite{sK03b}.


\subsection*{Acknowledgements}
C.D. is grateful to MIUR and the Alexander von Humboldt Stiftung for financial
support.  
S.K. acknowledges gratefully financial support from the
 \Name{Ev. Studienwerk Villigst} and helpful discussions with K.-H.\
Rehren (G\"ottingen).

 

\section*{Appendix}

The following technical lemma is crucial for the construction in section
\ref{sec:constimpl}. 

\label{apdiff}
\begin{lemma}
  \label{locmap}

Let $\Geb{U}_\varepsilon\subset \Diff_+(\Seins)$ ,$\varepsilon>0$
denote the neighbourhood of the identity

\begin{equation}\label{neighbor}
\{ \phi\in \Diff_{+}(\Seins)\, ; \sup_{z\in\Seins} { |\phi(z)-z| }<
\varepsilon \inf_{z\in\Seins} {|\phi(z)'|}  \}
\end{equation}   
Let $\{I_i\}_{i\in\dopp{Z}_m}$ be a finite covering  of
  the circle by proper intervals. Then for  $\varepsilon$ sufficiently 
  small  there exist continuous 
localising maps $\Xi_i: \Geb{U}_\varepsilon \longrightarrow
\Diff_{{I}_i}(\Seins)$ with the following features: 
$$
\phi = \prod_{i=1}^m \Xi_i (\phi)\, , \quad \Xi_i (id) = id\,.
$$
\end{lemma}

\begin{pf}
We look at the equivalent formulation in terms of periodic diffeomorphisms of
the real axis: 
$\varphi\in C^\infty(\dopp{R})$, $\varphi'(x)>0$,
$\varphi(x+2\pi)=\varphi(x) + 2\pi$.  The analogue of
$\varphi$ in $\Diff_+(\Seins)$ is denoted by $\check{\varphi}$. The preimage
of an interval $I\Subset\Seins$ 
under the covering projection $\symb{p}$ will be called $\hat{I}$. We choose a
smooth partition $\mu$ of unity on 
$\Seins$ satisfying $1\geq\mu_i\geq 0$,
$supp(\mu_i)\subset I_i$. On the covering space we define
$\lambda_i(x) := \mu_i(\symb{p}(x))$. 

Defining 
$  \psi_k[\varphi](x) := x + \sum_{i=1}^k \lambda_i(x)(\varphi(x)-x)$,
$ k=0,\ldots,m$,  
we have:
\begin{eqnarray}
  \Psi_k[\varphi]'(x)
&=&(1-\sum_{i=1}^k\lambda_i(x)) + \sum_{i=1}^k\lambda_i(x)\varphi'(x) + 
\sum_{i=1}^k\lambda_i'(x) (\varphi(x)-x)\nonumber\\
&\geqslant& \inf_{\xi\in\dopp{R}}(\min\{1,\varphi'(x)\}) - \sup_{\xi\in
  \dopp{R}}(\sum_{i=1}^m\betrag{\lambda_1'(\xi)}) \,\cdot\, \sup_{\xi\in
  \dopp{R}} |\varphi(\xi)-\xi| \label{eq:locbound}
\end{eqnarray}
For a  periodic diffeomorphism $ \varphi $ we have inf $ \varphi'(x)\leq 1$. With
\begin{center}
$\displaystyle \varepsilon^{-1}>\sup_{\xi\in\dopp{R}}\,\sum_{k=1}^m|
\lambda_k'(\xi)|\,$,
\end{center}
(\ref{neighbor}), (\ref{eq:locbound}), imply
\begin{center}
$\displaystyle \Psi_k[\varphi(x)]'\geq \Big(\inf \varphi'(x)\Big)\Big( 1-
\varepsilon \sup_{\xi\in\dopp{R}}\,\sum_{i=1}^k|\lambda_k'(\xi)|\Big)> 0\,$,
\end{center}
hence $\Psi_k[\varphi]$ is a
periodic diffeomorphism.

Moreover $ \Psi_k[\varphi]$ satisfies the estimate    $|\Psi_k[\varphi](x)-x|\leq
|\varphi(x)-x| < \varepsilon$. We can now choose $\varepsilon$ small 
enough , such that $|\chi(x)-x| < \varepsilon$ , for all $x$, 
$\chi\in\Diff_+(\Seins)$, implies $\chi(supp 
\lambda_k)\subset I_k$ .Then 
 $ \Psi_{k-1}[\varphi]^{-1} (I_k')\subset (supp \lambda_k)'$.  Moreover,
$\Psi_{k-1}[\varphi]$ and $\Psi_{k}[\varphi]$ coincide on $(supp \lambda_k)'$,
hence $\Xi_k(\varphi):= \Psi_{k}[\varphi]\circ \Psi_{k-1}[\varphi]^{-1}$
defines a periodic diffeomorphism, whose counterpart in $\Diff_+(\Seins)$,
namely $\check{\Xi}_k(\varphi)$, is localised in $I_k$. We define the {\em
  localising maps} on $\Geb{U}_{\varepsilon}$ as follows:
$\Xi_k(\check{\varphi}):= 
\check{\Xi}_k(\varphi)$.  

Continuous dependence of  $\Xi_k(\varphi)$ on $\varphi$ is obvious,
$\Psi_k[id]=id$ yields
$\Xi_k(id)=id$. Finally, with $\Psi_{m}[\varphi]= \varphi$ and
$\Psi_{0}[\varphi]= id$:
\begin{displaymath}
  \prod_{k=1}^m \Xi_k(\varphi) =
  \Psi_{m}[\varphi]\circ\Psi_{m-1}[\varphi]^{-1}\circ
  \Psi_{m-1}[\varphi]\circ\ldots\circ \Psi_{0}[\varphi]^{-1}
= \varphi\circ id
\end{displaymath}  
\end{pf}



\end{document}